%% file: corr.tex
\documentclass{article}

\usepackage{epsfig}
\usepackage{named}
\usepackage{amsmath,theorem}

\usepackage{xspace}
\usepackage{amssymb}
\usepackage{euscript}

\newfont{\mymathtt}{cmtt10 scaled 1095}

\newcommand{\alc}{%                             ALC
  \ensuremath{\mathcal{ALC}}\xspace}
\newcommand{\alci}{\si}
\newcommand{\si}{%
  \ensuremath{\mathcal{SI}}\xspace}
\newcommand{\siq}{%
  \ensuremath{\mathcal{SIQ}}\xspace}

\newcommand{\shi}{%
   \ensuremath{\mathcal{SHI}}\xspace}

\newcommand{\shiq}{\ensuremath{\mathcal{SHIQ}}\xspace}

\newcommand{\alcRtrans}{%\s}
        \ensuremath{\alc_{R^{+}}}\xspace}

\newcommand{\s}{%                                  ALC_R+
        \ensuremath{\mathcal{S}}\xspace}

\newcommand{\shif}{%                               ALCHFI_R+
  \ensuremath{\mathcal{SHIF}}\xspace}

\renewcommand{\sin}{%
  \ensuremath{\mathcal{SIN}}\xspace}               %ALCNI_R+

\newcommand{\shin}{%
  \ensuremath{\mathcal{SHIN}}\xspace} 

\newcommand{\Fact}{FaCT\xspace}%                FaCT
\newcommand{\PDL}%                              PDL
        {PDL\xspace}
\newcommand{\CPDL}%                             converse-PDL
        {\emph{converse}-PDL\xspace}

\newfont{\bigmathxx}{cmsy10 scaled 1440}
\newfont{\smallmathxx}{cmsy10 scaled 720}

\newcommand{\bigsqcap}{\mathop{\mathop{\mbox{\bigmathxx\symbol{117}}}}\limits}

\newcommand{\I}{%                               I (caligraphic)
        \ensuremath{\mathcal{I}}\xspace}

\newcommand{\Tree}{%                            Tree
        \ensuremath{\mathbf{T}}\xspace}
\newcommand{\ifunc}{%                           ^I (caligraphic)
        \ensuremath{^\mathcal{I}}\xspace}
\newcommand{\deltai}{%                          Delta^I
        \ensuremath{\Delta\ifunc}\xspace}

\newcommand{\Pspace}{\textsc{Pspace}}

\newcommand{\Term}{%                            Terminology
        \ensuremath{\mathcal{T}}\xspace}

\newcommand{\Concepts}{\ensuremath{\mathbf{C}}\xspace}
\newcommand{\Roles}{\ensuremath{\mathbf{R}}\xspace}
\newcommand{\Rplus}{\ensuremath{\mathbf{R}_+}\xspace}

\newcommand{\Lab}{\ensuremath{\EuScript{L}}\xspace}
\newcommand{\BLab}{\ensuremath{\EuScript{B}}\xspace}
\newcommand{\Edges}{\ensuremath{\EuScript{E}}\xspace}
\newcommand{\Inv}{\mathop{\mathsf{Inv}}}
\newcommand{\Tr}{\mathop{\mathsf{Trans}}}

\newcommand{\K}%                                Modal K
        {\ensuremath{\mathbf{K}}\xspace}
\newcommand{\KT}%                               Modal KT
        {\ensuremath{\mathbf{KT}}\xspace}
\newcommand{\Kfour}%                            Modal K4
        {\ensuremath{\mathbf{K4}}\xspace}
\newcommand{\Sfour}%                            Modal S4
        {\ensuremath{\mathbf{S4}}\xspace}
\newcommand{\Km}%                               Modal Km
        {\ensuremath{\K_{(\mathbf{m})}}\xspace}
\newcommand{\KTm}%                              Modal KTm
        {\ensuremath{\KT_{(\mathbf{m})}}\xspace}
\newcommand{\Kfourm}%                           Modal K4m
        {\ensuremath{\mathbf{K4}_{(\mathbf{m})}}\xspace}
\newcommand{\Sfourm}%                           Modal S4m
        {\ensuremath{\mathbf{S4}_{(\mathbf{m})}}\xspace}

\newcommand{\Tail}{\mathop{\mathsf{Tail}}}

\newcommand{\some}[2]{%
        \ensuremath{\exists #1 . #2}}
\newcommand{\all}[2]{%
        \ensuremath{\forall #1 . #2}}
\newcommand{\atleast}[2]{%
        \ensuremath{\mbox{$\geqslant$}#1 #2}}
\newcommand{\atmost}[2]{%
        \ensuremath{\mbox{$\leqslant$}#1 #2}}
\newcommand{\atleastq}[3]{%
        \ensuremath{\mbox{$\geqslant$}#1 #2 . #3}}
\newcommand{\atmostq}[3]{%
        \ensuremath{\mbox{$\leqslant$}#1 #2 . #3}}

\newcommand{\tuple}[2]{%                                binary tuple
        \ensuremath{\langle #1 , #2 \rangle}}

\newcommand{\Card}[1]{%                                 set cardinality
        \ensuremath{\vert#1\vert}}

\newcommand{\N}{\mathbb{N}}

\newcommand{\sss}{{\mathrel{\kern.25em{\sqsubseteq}\kern-.5em \mbox{{\scriptsize *}}\kern.25em}}}
\newcommand{\R}{\ensuremath{\mathcal{R}}\xspace}

\newcommand{\qnrleq}[3]{\ensuremath{(\leqslant #1 \; #2 \; #3)}}
\newcommand{\qnrgeq}[3]{\ensuremath{(\geqslant #1 \; #2 \; #3)}}

\newcommand{\clos}{\textit{clos}}
\newcommand{\nneg}{\ensuremath{\mathord{\sim}}}
\newcommand{\ndoteq}{\ensuremath{\mathrel{\not\doteq}}}
\newcommand{\Paths}{\ensuremath{\mathsf{Paths}}\xspace}

  \theoremstyle{break}
  \theoremheaderfont{\bf\boldmath}
  \newtheorem{satz}{Satz}[section]
  \newtheorem{theorem}[satz]{Theorem}
  \newtheorem{definition}[satz]{Definition}
  \theoremstyle{plain}
  \newtheorem{lemma}[satz]{Lemma}

\newcommand{\qed}{{\unskip\nobreak\hfil\penalty50
   \hskip2em\hbox{}\nobreak\hfil
   \qedd
   \parfillskip=0pt \finalhyphendemerits=0
    \medskip\goodbreak\noindent}}
\newcommand{\qedd}{\vrule height4pt width 4pt depth0pt}

\title{Practical Reasoning for Expressive Description Logics\thanks{This paper
    appeared in the Proceedings of the 6th International Conference on Logic
    for Programming and Automated Reasoning (LPAR'99), number 1704 Lecture
    Notes in Artificial Intelligence, pages 161-180. Springer-Verlag,
    September 1999.}}

\author{{\bf Ian Horrocks}\\
  Department of Computer Science, University of Manchester\\
  \texttt{horrocks@cs.man.ac.uk} \and 
  {\bf Ulrike Sattler}\\
  LuFG Theoretical Computer Science, RWTH Aachen\\
  \texttt{sattler@informatik.rwth-aachen.de} \and 
  {\bf Stephan Tobies}\\
  LuFG Theoretical Computer Science, RWTH Aachen\\
  \texttt{tobies@informatik.rwth-aachen.de}}

\date{}

\begin{document}
\maketitle

\begin{abstract}
  Description Logics (DLs) are a family of knowledge representation formalisms
  mainly characterised by constructors to build complex concepts and roles from
  atomic ones. Expressive role constructors are important in many applications,
  but can be computationally problematical.  We present an algorithm that
  decides satisfiability of the DL \alc extended with transitive and inverse
  roles, role hierarchies, and qualifying number restrictions. Early experiments
  indicate that this algorithm is well-suited for implementation.  Additionally,
  we show that \alc extended with just transitive and inverse roles is still in
  \textsc{PSpace}. Finally, we investigate the limits of decidability for this
  family of DLs.
\end{abstract}

\section{Motivation}\label{sec:motivat}

Description Logics (DLs) are a well-known family of knowledge representation
formalisms~\cite{DLNS96}. They are based on the notion of concepts (unary
predicates, classes) and roles (binary relations), and are mainly characterised
by constructors that allow complex concepts and roles to be built from atomic
ones.  Sound and complete algorithms for the interesting inference problems such
as subsumption and satisfiability of concepts are known for a wide variety of
DLs~\cite{SSSm91,DLNN91,Sat95ashort,GiLe96,CaldGL-IJCAI-99}.

To be used in a specific application, the expressivity of the DL must be sufficient
to describe relevant properties of objects in the application domain.
For example, transitive roles (e.g. ``ancestor'') and inverse roles (e.g.
``successor''/``predecessor'') play an important r\^ole not only in the adequate
representation of complex, aggregated objects~\cite{HoSat98c}, but also for
reasoning with conceptual data models~\cite{CaLN94}. Moreover, reasoning with
respect to cyclic definitions is crucial for applying DLs to reasoning with
database schemata \cite{CaldGL-PODS-98}.

The relevant inference problems for (extensions of) DLs that allow for
transitive and inverse roles are known to be decidable~\cite{GiLe96}, and
appropriate inference algorithms have been described~\cite{DeGiaMass98}, but
their high degree of non-determinism appears to prohibit their use in realistic
applications.  This is mainly due to the fact that these algorithms can handle
not just transitive roles but also the transitive closure of roles. It has been
shown~\cite{Sat95ashort} that restricting a DL to transitive roles can lead to a
lower complexity, and that transitive roles (even when combined with role
hierarchies) allow for algorithms that behave quite well in realistic
applications~\cite{Horrocks98c}.  However, it remained to show that this is
still true when inverse roles and qualifying number restrictions are also present.

This paper extends our understanding of these issues in several directions.
Firstly, we present an algorithm that decides satisfiability of \alc
\cite{SSSm91} (which can be seen as a notational variant of the multi modal logic
$\mathsf{K}_m$) extended with transitive and inverse roles, role hierarchies,
and qualifying number restrictions, i.e., concepts of the form $\qnrgeq 3
{hasChild} {Female}$ that allow the description of objects by restricting the number of
objects of a given type they are related to via a certain role.  The algorithm
can also be used for checking satisfiability and subsumption with respect to
general concept inclusion axioms (and thus cyclic definitions) because these
axioms can be ``internalised''.  The absence of
transitive closure leads to a % considerably
lower degree of non-determinism, and experiments indicate that the algorithm is
well-suited for implementation.

Secondly, we show that \alc extended with both transitive \emph{and}
inverse roles is still in \Pspace. The algorithm used to prove this
rather surprising result introduces an enhanced \emph{blocking}
technique. In general, blocking is used to ensure termination of the
algorithm in cases where it would otherwise be stuck in a loop. The
enhanced blocking technique allows such cases to be detected earlier
and should provide useful efficiency gains in implementations of this
and more expressive DLs.

Finally, we investigate the limits of decidability for this family of DLs,
showing that relaxing the constraints placed on the kinds of roles allowed in number
restrictions leads to the undecidability of all inference problems.

Due to a lack of space we can only present selected proofs.  For full details
please refer to \cite{Horrocks98k,HoSatTo99}.

\section{Preliminaries}

In this section, we present the syntax and semantics of the various DLs that are
investigated in subsequent sections. This includes the definition of inference
problems (concept subsumption and satisfiability, and both of these problems
with respect to terminologies) and how they are interrelated.

The logics we will discuss are all based on an extension of the well known DL
\alc~\cite{SSSm91} to include transitively closed primitive
roles~\cite{Sat95ashort}; we will call this logic \s due to its relationship
with the proposition (multi) modal logic \Sfourm~\cite{Schi91}.\footnote{The
  logic \s has previously been called \alcRtrans, but this becomes too cumbersome
  when adding letters to represent additional features.} This basic DL is then
extended in a variety of ways---see Figure~\ref{fig:Sfamily} for an overview.

\begin{definition}\label{syntax+semantics}
  Let $\Concepts$ be a set of \emph{concept names} and \Roles a set of
  \emph{role names} with transitive role names $\Rplus\subseteq \Roles$.
  The set of \si -{\em roles\/} is $\Roles \cup \{R^-\mid R\in\Roles \}$.
  The set of \si-\emph{concepts} is the smallest set such that
                                %
                                %  \begin{enumerate}
                                %  \item
  every concept name is a concept, and,
                                %  \item
  if $C$ and $D$ are concepts and $R$ is an \alci -role, then $(C\sqcap D)$,
  $(C\sqcup D)$, $(\neg C)$, $(\forall R.C)$, and $(\exists R.C)$ are also
  concepts.
                                %  \end{enumerate}

  To avoid considering roles such as $R^{--}$, we define a function $\Inv$ on
  roles such that $\Inv(R) = R^-$ if $R$ is a role name, and $\Inv(R) = S$ if
  $R=S^-$.  We also define a function $\Tr$ which returns $\mathrm{true}$ iff
  $R$ is a transitive role. More precisely, $\Tr(R) = \mathrm{true}$ iff $R\in
  \Rplus$ or $\Inv(R)\in\Rplus$.

  \shi is obtained from \si by allowing, additionally, for a set of \emph{role
  inclusion axioms} of the form $R\sqsubseteq S$, where $R$ and $S$ are two
  roles, each of which can be inverse. For a set of role inclusion axioms \R,
  $$\R^+:=(\R\cup\{ \Inv(R)\sqsubseteq \Inv(S)\mid R\sqsubseteq S \in \R\}
  ,\sss)$$
  is called a {\em role hierarchy}, where $\sss$ is the transitive-reflexive
  closure of $\sqsubseteq$ over $\R\cup\{ \Inv(R)\sqsubseteq \Inv(S)\mid
  R\sqsubseteq S \in \R\}$.

  \shiq is obtained from \shi by allowing, additionally, for \emph{qualifying
    number restrictions}, i.e., for concepts of the form $\qnrgeq n R C$ and
  $\qnrleq n R C$, where $R$ is a \emph{simple} (possibly inverse) role and
  $n$ is a non-negative integer. A role is called \emph{simple} iff it is
  neither transitive nor has transitive sub-roles.

  \shin is the restriction of \shiq where qualifying number restrictions may
  only be of the form $\qnrgeq n R \top$ and $\qnrleq n R \top$. In this case, we
  omit the symbol $\top$ and write $(\geqslant n \; R)$ and $(\leqslant n \; R)$
  instead.

  An {\em interpretation\/} $\I = (\Delta^\I,\cdot^\I)$ consists of a set
  $\Delta^\I$, called the {\em domain\/} of $\I$, and a \emph{valuation}
  $\cdot^\I$ which maps every concept to a subset of $\Delta^\I$ and every role
  to a subset of $\Delta^\I\times\Delta^\I$ such that, for all concepts $C$,
  $D$, roles $R$, $S$, and non-negative integers $n$, the properties in
  Figure~\ref{fig:Sfamily} are satisfied, where $\sharp M$ denotes the
  cardinality of a set $M$. An interpretation satisfies a role hierarchy $\R^+$
  iff $R^\I \subseteq S^\I$ for each $R \sss S \in \R^+$; we denote this
  fact by $\I \models \R^+$ and say that $\I$ is a model of $\R^+$.

  A concept $C$ is called {\em satisfiable\/} with respect to a role hierarchy
  $\R^+$ iff there is some interpretation $\I$ such that $\I \models \R^+$ and
  $C^\I \neq \emptyset$.  Such an interpretation is called a {\em model of\/}
  $C$ w.r.t. $\R^+$. A concept $D$ {\em subsumes\/} a concept $C$ w.r.t. $\R^+$
  (written $C \sqsubseteq_{\R^+} D$) iff $C^\I \subseteq D^\I$ holds for each
  model $\I$ of $\R^+$.  For an interpretation $\I$, an individual $x \in
  \Delta^\I$ is called an {\em instance} of a concept $C$ iff $x\in C^\I$.
\end{definition}
All DLs considered here are closed under negation, hence subsumption and
(un)satisfiability w.r.t. role hierarchies can be reduced to each other: $C
\sqsubseteq_{\R^+} D$ iff $C\sqcap \neg D$ is unsatisfiable w.r.t.  $\R^+$, and
$C$ is unsatisfiable w.r.t. $\R^+$ iff $C\sqsubseteq_{\R^+} A\sqcap \neg A$ for
some concept name $A$.

\begin{figure}[t!]
\begin{center}
{\setlength{\arraycolsep}{0em}
\begin{tabular}{|@{$\,$}l@{$\,$}|@{$\,$}c@{$\,$}|@{$\,$}c@{$\,$}|@{$\,$}c@{$\,$}|}
\hline
Construct Name & Syntax & Semantics & \\
\hline
atomic concept
 & $A$ & $A\ifunc \subseteq \deltai$ & \\
universal concept
 & $\top$ & $\top\ifunc = \deltai$ & \\
 atomic role
 & $R$ & $R\ifunc \subseteq \deltai \times \deltai$ & \\
transitive role
 & $R \in \Rplus$ & $R\ifunc = (R\ifunc)^+$ & \\
conjunction
 & $C \sqcap D$ & $C\ifunc \cap D\ifunc$ & \\
disjunction
 & $C \sqcup D$ & $C\ifunc \cup D\ifunc$ & $\mathcal{S}$ \\
negation
 & $\neg C$ & $\deltai\setminus C\ifunc$ & \\
exists restriction
 & \some{R}{C}
 & $\{x \mid \exists y.\tuple{x}{y} \in R\ifunc \mbox{ and }y \in C\ifunc\}$ &\\
value restriction
 &\all{R}{C}
 & $\{x \mid \forall y.\tuple{x}{y} \in R\ifunc \mbox{ implies }y \in
 C\ifunc\}$ & \\
\hline
role hierarchy
 & $R \sqsubseteq S$ & $R\ifunc \subseteq S\ifunc$ & $\mathcal{H}$ \\
\hline
inverse role
 & $R^-$ & $\{\tuple{x}{y} \mid \tuple{y}{x} \in R\ifunc\} $ & $\mathcal{I}$ \\
\hline
$\begin{array}{@{}l@{}}
 \mbox{number}\\
 \mbox{restrictions}
 \end{array}$
 & $\begin{array}{c}
    \atleast{n}{R} \\
    \atmost{n}{R}
    \end{array}$
 & $\begin{array}{l}
   \{x \mid \sharp \{y.\tuple{x}{y} \in R\ifunc\} \geqslant n\} \\
   \{x \mid \sharp \{y.\tuple{x}{y} \in R\ifunc\} \leqslant n\}
    \end{array}$
 & $\mathcal{N}$ \\
\hline
$\begin{array}{@{}l@{}}
 \mbox{qualifying } \\
 \mbox{number}\\
 \mbox{restrictions}
 \end{array}$
 & $\begin{array}{c}
    \atleastq{n}{R}{C} \\
    \atmostq{n}{R}{C}
    \end{array}$
 & $\begin{array}{l}
    \{x \mid \sharp \{y.\tuple{x}{y} \in R\ifunc \mbox{ and } y \in C\ifunc\} \geqslant n\} \\
    \{x \mid \sharp \{y.\tuple{x}{y} \in R\ifunc \mbox{ and }  y \in C\ifunc\} \leqslant n\}
    \end{array}$
 & $\mathcal{Q}$ \\
\hline
\end{tabular}}
\end{center}
\caption{Syntax and semantics of the \si family of DLs}\label{fig:Sfamily}
\end{figure}

In \cite{Baad90c,Schi91,BBNNS93}, the \emph{internalisation} of terminological
axioms is introduced, a technique that reduces reasoning with respect to a
(possibly cyclic) terminology to satisfiability of concepts.  In
\cite{Horrocks98c}, we saw how role hierarchies can be used for this reduction.
In the presence of inverse roles, this reduction must be slightly modified.

\begin{definition}

  A \emph{terminology} \Term is a finite set of \emph{general concept inclusion
    axioms}, $\Term=\{C_1\sqsubseteq D_1,\ldots, C_n\sqsubseteq D_n\}$, where
  $C_i,D_i$ are arbitrary \shiq-concepts. An interpretation $\I$ is said to be a
  \emph{model} of \Term iff $C_i\ifunc \subseteq D_i\ifunc$ holds for all $C_i
  \sqsubseteq D_i\in \Term$.  $C$ is \emph{satisfiable} with respect to \Term
  iff there is a model $\I$ of \Term with $C\ifunc \not = \emptyset$.  Finally,
  $D$ \emph{subsumes} $C$ with respect to \Term iff for each model $\I$ of \Term
  we have $C\ifunc \subseteq D\ifunc$.
\end{definition}

The following Lemma shows how general concept inclusion axioms can be
\emph{internalised} using a ``universal'' role $U$, that is, a
transitive super-role of all roles occurring in \Term and their respective inverses.

\begin{lemma}\label{lemma:terminologies}
  Let \Term be a terminology, $\R$ a set of role inclusion axioms
  % , $\R^+$ the  associated role hierarchy,
  and $C,D$ \shiq-concepts and let
  $$C_\Term:= \bigsqcap_{C_i\sqsubseteq D_i \in\Term}\neg C_i\sqcup D_i.$$
  \noindent  Let $U$ be a transitive role that does not occur in $\Term,C,D$, or $\R$. We
  set
  $$\R_U := \R \cup \{ R\sqsubseteq U, \Inv(R)\sqsubseteq U \mid \text{$R$
  occurs in $\Term,C,D$, or $\R$} \}.$$
  Then $C$ is satisfiable w.r.t. \Term and $\R^+$ iff
  $C\sqcap C_\Term \sqcap \all{U}{C_\Term}$
  is satisfiable w.r.t. $\R_U^+$. Moreover, $D$ subsumes $C$ with respect to
  \Term and $\R^+$ iff
  $C\sqcap\neg D\sqcap C_\Term \sqcap \all{U}{C_\Term}$
  is unsatisfiable w.r.t. $\R_U^+$.
\end{lemma}

The proof of Lemma~\ref{lemma:terminologies} is similar to the ones that can be
found in \cite{Schi91,Baad90c}. Most importantly, it must be shown that, (a) if
a \shiq-concept $C$ is satisfiable with respect to a terminology \Term and a
role hierarchy $\R^+$, then $C,\Term$ have a \emph{connected} model, and (b) if
$y$ is reachable from $x$ via a role path (possibly involving inverse roles),
then $\tuple{x}{y} \in U\ifunc$. These are easy consequences of the semantics
and the definition of $U$.

\begin{theorem}\label{theorem:internal}
  Satisfiability and subsumption of \shiq-concepts (resp.\ \shi-con\-cepts)
  w.r.t. terminologies and role hierarchies are polynomially reducible to
  (un)sat\-is\-fi\-abil\-i\-ty of \shiq-concepts (resp.\ \shi-concepts) w.r.t.
  role hierarchies.
\end{theorem}

\section{Reasoning for \si Logics} \label{sec:alchfi}

\newcommand{\Uhf}{} \newcommand{\path}[1]{[#1]}

In this section, we present two tableaux algorithms: the first decides
satisfiability of \shiq-concepts, and can be used for all \shiq reasoning
problems (see Theorem~\ref{theorem:internal}); the second decides satisfiability
(and hence subsumption) of \si-concepts in \Pspace. Please note that \shin (and
hence \shiq) no longer has the finite model property: for example, the following
concept, where $R$ is a transitive super-role of $F$, is satisfiable, but each
of its models has an infinite domain.
$$\neg C \sqcap \some{F^-}{(C \sqcap \atmost{1}{F})} \sqcap
\all{R^-}{(\some{F^-}{(C\sqcap \atmost{1}{F} )}) }$$

This concept requires the existence of an infinite $F^-$-path, where the first
element on the path satisfies $\neg C$ while all other elements satisfy $C
\sqcap \atmost{1}{F}$. This path cannot collapse into a cycle: (a) it
cannot return to the first element because this element cannot satisfy both $C$
and $\neg C$; (b) it cannot return to any subsequent element on the path because
then this node would not satisfy $\atmost{1}{F}$.

The correctness of the algorithms we are presenting can be proved by showing
that they create a \emph{tableau} for a concept iff it is satisfiable.
For ease of construction, we assume all concepts to be in \emph{negation normal
form} (NNF), that is, negation occurs only in front of concept names.  Any
\shiq-concept can easily be transformed to an equivalent one in NNF by pushing
negations inwards~\cite{HoNS90}; with $\nneg C$ we denote the NNF of $\neg C$.
For a concept $C$ in NNF we define $\clos(C)$ as the smallest set of concepts
that contains $C$ and is closed under subconcepts and $\nneg$.  Please note that
size of $\clos(C)$ is linearly bounded by the size of $C$.

\begin{definition} \label{def:alchr2}
  Let $D$ be a \shiq-concept in NNF, $\R^+$ a role hierarchy, and $\Roles_D$ the
  set of roles occurring in $D$ and $\R^+$ together with their inverses. Then $T
  = (\mathbf{S},\Lab,\Edges)$ is a \emph{tableau} for $D$ w.r.t. $\R^+$ iff
  $\mathbf{S}$ is a set of individuals, $\Lab:\mathbf{S} \rightarrow
  2^{\clos(D)}$ maps each
  individual to a set of concepts, %  which is a subset of $\textit{sub}(D)$,
  $\Edges:\Roles_D \rightarrow 2^{\mathbf{S} \times \mathbf{S}}$ maps each role
  to a set of pairs of individuals, and there is some individual $s \in
  \mathbf{S}$ such that $D \in \Lab(s)$. Furthermore, for all $s,t \in
  \mathbf{S}$, $C, C_1, C_2\in \clos(D)$, and $R,S \in \Roles_D$, it holds that:
  \begin{enumerate}\setlength{\itemsep}{0ex}
  \item if $C \in \Lab(s)$, then $\neg C \notin \Lab(s)$,
  \item if $C_1 \sqcap C_2 \in \Lab(s)$, then $C_1 \in \Lab(s)$ and $C_2 \in
    \Lab(s)$,
  \item if $C_1 \sqcup C_2 \in \Lab(s)$, then $C_1 \in \Lab(s)$ or $C_2 \in
    \Lab(s)$,
  \item if $\all{S}{C} \in \Lab(s)$ and $\tuple{s}{t} \in \Edges(S)$, then $C \in
    \Lab(t)$,
  \item if $\some{S}{C} \in \Lab(s)$, then there is some $t \in \mathbf{S}$ such
    that $\tuple{s}{t} \in \Edges(S)$ and $C \in \Lab(t)$,
  \item if $\all{S}{C} \in \Lab(s)$ and $\tuple{s}{t} \in \Edges(R)$ for some $R
    \sss S$ with $\Tr(R)$, then $\all{R}{C} \in \Lab(t)$,
  \item $\tuple{x}{y}\in \Edges(R)$ iff $\tuple{y}{x}\in \Edges(\Inv(R))$,
  \item if $\tuple s t \in \Edges(R)$ and $R \sss S$, then $\tuple s t \in
    \Edges(S)$,
  \item if $\qnrleq n S C \in \Lab(s)$, then $\sharp S^T(s,C) \leqslant n$,
  \item if $\qnrgeq n S C \in \Lab(s)$, then $\sharp S^T(s,C) \geqslant n$,
  \item if $(\bowtie \; n \; S \; C) \in \Lab(s)$ and $\tuple s t \in \Edges(S)$
    then $C \in \Lab(t)$ or $\nneg C \in \Lab(t)$,
  \end{enumerate}
  where we use $\bowtie$ as a placeholder for both $\leqslant$ and $\geqslant$ and we
  define
  \[
  S^T(s,C) := \{ t \in \mathbf{S} \mid \tuple s t \in \Edges(S) \
  \text{and} \ C \in \Lab(t) \} .
  \]

  Tableaux for \si-concepts are defined analogously and must satisfy Properties
  1-7, where, due to the absence of a role hierarchy, $\sss$ is the identity.
\end{definition}

Due to the close relationship between models and tableaux, the following lemma
can be easily proved by induction.  As a consequence, an
algorithm that constructs (if possible) a tableau for an input concept is a
decision procedure for satisfiability of concepts.

\begin{lemma}\label{lemma:shin-tabl}
  A \shiq-concept (resp.\ \si-concept) $D$ is satisfiable w.r.t. a role hierarchy
  $\R^+$ iff $D$ has a tableau w.r.t. $\R^+$.
\end{lemma}

\subsection{Reasoning in \shiq}\label{sec:shin-algo}

In the following, we give an algorithm that, given a \shiq-concept $D$, decides
the existence of a tableaux for $D$. We implicitly assume an arbitrary but fixed
role hierarchy $\R^+$.  The tableaux algorithm works on a finite {\em completion
tree} (a tree some of whose nodes correspond to individuals in the tableau, each
node being labelled with a set of \shiq-concepts), and employs a \emph{blocking}
technique \cite{HoSat98c} to guarantee termination: If a path contains two pairs
of successive nodes that have pair-wise identical label and whose connecting
edges have identical labels, then the path beyond the second pair is no longer
expanded, it is said to be blocked. Blocked paths can be ``unravelled'' to
construct an infinite tableau.  The identical labels make sure that copies of
the first pair and their descendants can be substituted for the second pair of
nodes and their respective descendants.

\begin{definition}\label{def:shin-algo}
  A \emph{completion tree} for a \shiq-concept $D$ is a tree where each node $x$
  of the tree is labelled with a set $\Lab(x) \subseteq \clos(D)$ and each edge
  $\tuple{x}{y}$ is labelled with a set $\Lab(\tuple{x}{y})$ of (possibly
  inverse) roles occurring in $\clos(D)$; explicit inequalities between nodes of
  the tree are recorded in a binary relation ${}\ndoteq{}$ that is implicitly
  assumed to be symmetric.

  Given a completion tree, a node $y$ is called an $R$-\emph{successor} of a
  node $x$ iff $y$ is a successor of $x$ and $S\in \Lab(\tuple{x}{y})$ for some
  $S$ with $S\sss R$.  A node $y$ is called an $R$\emph{-neighbour} of $x$ iff
  $y$ is an $R$-successor of $x$, or if $x$ is an $\Inv(R)$-successor of $y$.
  Predecessors and ancestors are defined as usual.

  A node is \emph{blocked} iff it is directly or indirectly blocked.  A node $x$
  is \emph{directly blocked} iff none of its ancestors are blocked, and it has
  ancestors $x'$, $y$ and $y'$ such that
  \begin{enumerate}\setlength{\itemsep}{0ex}
    \item $x$ is a successor of $x'$ and $y$ is a successor of $y'$ \emph{and}
    \item $\Lab(x) = \Lab(y)$ and $\Lab(x') = \Lab(y')$ \emph{and}
    \item $\Lab(\tuple{x'}{x}) = \Lab(\tuple{y'}{y})$.
  \end{enumerate}
  In this case we will say that $y$ \emph{blocks} $x$. Since this blocking
  technique involves
  pairs of nodes, it is called \emph{pair-wise} blocking.

  A node $y$ is \emph{indirectly blocked} iff one of its ancestors is blocked,
  or %a node $y$ will also be taken to be indirectly blocked if
  it is a successor of a node $x$ and $\Lab(\tuple{x}{y}) = \emptyset$; the
  latter condition avoids wasted expansions after an application of the
  $\leqslant$-rule.

  For a node $x$, $\Lab(x)$ is said to contain a {\em clash} iff $\{A,\neg A\}
  \subseteq \Lab(x)$ or if, for some concept $C$, some role $S$, and some $n
  \in \N$: $\qnrleq n S C \in \Lab(x)$ and there are $n+1$ $S$-neighbours
  $y_0,\dots,y_n$ of $x$ such that $C \in \Lab(y_i)$ and $y_i \ndoteq y_j$ for
  all $0 \leq i < j \leq n$.  A completion tree is called {\em clash-free} iff
  none of its nodes contains a clash; it is called {\em complete} iff none of
  the expansion rules in Figure~\ref{table:alchfi} is applicable.

  For a \shiq-concept $D$, the algorithm starts with a completion tree
  consisting of a single node $x$ with $\Lab (x)=\{D\}$ and ${\ndoteq} =
  \emptyset$. It applies the expansion rules in Figure~\ref{table:alchfi},
  stopping when a clash occurs, and answers ``$D$ is satisfiable'' iff the
  completion rules can be applied in such a way that they yield a complete and
  clash-free completion tree.
\end{definition}

\begin{figure}[t!]
  \begin{center}
     \small{
    \framebox[\textwidth]{\begin{tabular}{@{ }l@{ }l@{ }l@{}}

        $\sqcap$-rule: & if \hfill 1. & $C_1 \sqcap C_2 \in \Lab(x)$, $x$ is not
        indirectly blocked, and \\
        & \hfill 2. & $\{C_1,C_2\} \not\subseteq \Lab(x)$ \\

        & then & $\Lab(x) \longrightarrow \Lab(x) \cup \{C_1,C_2\}$ \\[1ex]

        $\sqcup$-rule: & if \hfill 1. & $C_1 \sqcup C_2 \in \Lab(x)$, $x$ is not
        indirectly blocked, and \\
        & \hfill 2. & $\{C_1,C_2\} \cap \Lab(x) = \emptyset$ \\

        & then & $\Lab(x) \longrightarrow \Lab(x) \cup \{C\}$
        for some $C\in \{C_1,C_2 \}$\\[1ex]

        $\exists$-rule: & if \hfill 1. & $\some{S}{C} \in \Lab(x)$, $x$ is not blocked, and \\
        & \hfill 2. & $x$ has no $S$-neighbour $y$ with $C\in \Lab(y)$, \\

        & then & create a new node  $y$  with $\Lab(\tuple{x}{y})=\{S\}$
        and $\Lab(y)=\{C\}$ \\[1ex]

        $\forall$-rule: & if \hfill 1. & $\all{S}{C} \in \Lab(x)$, $x$ is not
        indirectly blocked, and \\
        & \hfill 2. &there is an $S$-neighbour $y$ of $x$ with $C \notin \Lab(y)$ \\

        & then & $\Lab(y) \longrightarrow \Lab(y) \cup \{C\}$ \\[1ex]

        $\forall_+$-rule: & if \hfill 1. & $\all{S}{C} \in \Lab(x)$,  $x$ is not
        indirectly blocked, and\\
        & \hfill 2. & there is some $R$ with  $\Tr(R)$ and $R\sss S$, \\
        & \hfill 3. &there is an  $R$-neighbour $y$ of $x$ with $\all{R}{C}
        \notin \Lab(y)$ \\
        & then & $\Lab(y) \longrightarrow \Lab(y) \cup \{\all{R}{C}\}$\\[1ex]

        \textit{choose}-rule: & if \hfill 1. & $(\bowtie \; n \; S \; C) \in \Lab(x)$,
        $x$ is not indirectly blocked, and\\
        & \hfill 2. & there is an $S$-neighbour $y$ of $x$ with $\{ C,\nneg C\}
        \cap  \Lab(y)= \emptyset$\\

        & then & $\Lab(y) \longrightarrow \Lab(y) \cup \{E\}$
        for some $E \in \{C,\nneg C \}$\\[1ex]

        $\geqslant$-rule: & if \hfill 1. & $\qnrgeq n S C \in \Lab(x)$, $x$ is not
        blocked, and\\
        & \hfill 2. & there are not $n$ $S$-neighbours  $y_1,\dots,y_n$ of $x$
        with \\
        && $C \in \Lab(y_i)$ and  $y_i \ndoteq y_j$ for $1 \leq i < j \leq n$\\

        & then & create $n$ new nodes $y_1,\dots,y_n$ with $\Lab(\tuple x {y_i}) =
        \{S\}$,\\
        & & $\Lab(y_i) = \{C\}$, and $y_i \ndoteq y_j$ for $1 \leq i < j \leq n$.\\[1ex]

        $\leqslant$-rule: & if \hfill 1. &  $\qnrleq n S C \in \Lab(x)$, $x$ is not indirectly
        blocked, and\\
        & \hfill 2. & $\sharp S^\Tree(x,C) > n$ and there are two $S$-neighbours
        $y,z$ of $x$ with \\
        & &  $C \in \Lab(y), C \in \Lab(z)$, $y$ is not an ancestor
        of $x$, and not $y \ndoteq z$\\
        &then & 1.\ $\Lab(z) \longrightarrow \Lab(z) \cup \Lab(y)$ and\\
        & & 2.\ \begin{tabular}[t]{rl}
          \multicolumn{2}{l}{if $z$ is an ancestor of $x$} \\
          then &
          $\begin{array}[t]{rcl}
            \Lab (\tuple{z}{x})
            &\longrightarrow&  \Lab (\tuple{z}{x}) \cup \Inv(\Lab (\tuple{x}{y}))
          \end{array}$\\
          else \hfill &
          $\begin{array}[t]{rcl}
            \Lab (\tuple{x}{z})
            &\longrightarrow&  \Lab (\tuple{x}{z}) \cup \Lab (\tuple{x}{y})
          \end{array}$
        \end{tabular}\\
        & & 3.\ $\Lab(\tuple{x}{y}) \longrightarrow \emptyset$ \\
        & & 4.\ Set $u \ndoteq z$ for all $u$ with $u \ndoteq y$
      \end{tabular}
      }}

    \caption{The complete tableaux expansion rules for
    \shiq}\label{table:alchfi}
    \vspace{-0.6cm}
  \end{center}
\end{figure}

  %NOTE - this isn't right - blocking may also necessary for logics for
  %logics that do have the FMP. We could say something about needing
  %the more sophisticated blocking strategy but in view of the space
  %constraints I suggest we just leave it out.
The soundness and completeness of the tableaux
algorithm is an immediate consequence of Lemmas~\ref{lemma:shin-tabl} and
\ref{lemma:shin-algo-correct}.

\begin{lemma} \label{lemma:shin-algo-correct}
  Let $D$ be an \shiq-concept.
  \begin{enumerate}
  \item The tableaux algorithm terminates when started with $D$.
  \item If the expansion rules can be applied to $D$ such
    that they yield a complete and clash-free completion tree, then $D$ has a
    tableau.
  \item If $D$ has a tableau, then the expansion rules can be applied to
    $D$ such that they yield a complete and clash-free completion
    tree.
  \end{enumerate}
\end{lemma}

The proof can be found in the appendix. Here, we will only discuss the intuition
behind the expansion rules and their correspondence to the constructors of
\shiq. Roughly speaking,\footnote{For the following considerations, we employ a
simpler view of the correspondence between completion trees and models, and need
not bother with the path construction mentioned above.}
the completion tree is a partial description of a model whose individuals
correspond to nodes, and whose interpretation of roles is taken from the edge
labels. Since the completion tree is a tree, this would not yield a correct
interpretation of transitive roles, and thus the interpretation of transitive
roles is built via the transitive closure of the relations induced by the
corresponding edge labels.

The $\sqcap$-, $\sqcup$-, $\exists$- and $\forall$-rules are the standard
tableaux rules for \alc or the propositional modal logic $\mathsf{K}_{m}$.  The
$\forall_+$-rule is the standard rule for $\alc_{R^+}$ or the propositional
modal logic $\mathsf{S4}_m$ extended to deal with role-hierarchies as follows.
Assume a situation that satisfies the precondition of the
$\forall_+$-rule, i.e., $\all{S}{C} \in \Lab(x)$, and there is an
$R$-neighbour $y$ of $x$ with $\Tr(R)$, $R\sss S$ and $\all{R}{C}
        \notin \Lab(y)$.
If $y$ has an $R$-successor $z$, then, due to the transitivity of $R$, $z$ is also
an $R$-successor of $x$. Since $R \sss S$, it is also an $S$-successor of $x$
and hence must satisfy $C$. This is ensured by adding $\all{R}{C}$ to $\Lab(z)$

The rules dealing with qualifying number restrictions work similarly to the
rules given in \cite{Baader96a}. For a concept $\qnrgeq n R C \in \Lab(x)$, the
$\geqslant$-rule generates $n$ $R$-successors $y_1,\dots,y_n$ of $x$ with $C \in
\Lab(y_i)$ . To prevent the $\leqslant$-rule from indentifying the new nodes, it
also sets $y_i \ndoteq y_j$ for each $1\leq i < j \leq n$ .  Conversely, if
$\qnrleq n R C \in \Lab(x)$ and $x$ has more than $n$ $R$-neighbours that are
labelled with $C$, then the $\leqslant$-rule chooses two of them that are not in
$\ndoteq$ and merges them, together with the edges connecting them with $x$.  The
definition of a clash takes care of the situation where the $\ndoteq$ relation
makes it impossible to merge any two $R$-neighbours of $x$, while the
\textit{choose}-rule ensures that all $R$-neighbours of $x$ are labelled with
either $C$ or $\nneg C$. Without this rule, the unsatisfiability of concepts like
$\qnrgeq 3 R A \sqcap \qnrleq 1 R B \sqcap \qnrleq 1 R {\neg B}$ would go
undetected. The relation $\ndoteq$ is used to prevent infinite sequences of rule
applications for contradicting number restrictions of the form $\qnrgeq n R C$
and $\qnrleq {(m)} R C$, with $n>m$.
Labelling edges with sets of roles allows a single node to be both an $R$ and
$S$-successor of $x$ even if $R$ and $S$ are not comparable with respect to
$\sss$.

The following theorem is an immediate consequence of
Lemma~\ref{lemma:shin-tabl} and \ref{lemma:shin-algo-correct}, and
Theorem~\ref{theorem:internal}.

\begin{theorem}\label{theorem:dec-shin}
  The tableaux algorithm is a decision procedure for the satisfiability and
  subsumption of \shiq-concepts with respect to terminologies.
\end{theorem}

\newcommand{\pspace}{\textsc{PSpace}\xspace}

\subsection{A \pspace-algorithm for \si}

To obtain a (worst-case) optimal algorithm for \si, the \shiq
algorithm is modified as follows.
(a) Since \si does not allow for qualifying number restrictions the
$\geqslant$-, $\leqslant$-, and \textit{choose}-rule can be omitted.
In the absence of the \textit{choose}-rule we may assume all concepts
appearing in labels to be in NNF from the (smaller) set of all
subconcepts of $D$ denoted by $\textit{sub}(D)$, and in the absence of
role hierarchies, edge labels can be restricted to roles (instead of
sets of roles).  Due to the absence of number restrictions the logic
still has the finite model property, and blocking no longer need
involve two pairs of nodes with identical labels, but only two nodes
with (originally) identical labels.
(b) To obtain a \pspace algorithm, we employ a refined blocking
strategy which further loosens this ``identity'' condition to a
``similarity'' condition. This is achieved by using a second label
$\BLab$ for each node. In the following, we will describe and motivate
this blocking technique; detailed proofs as well as an extension of
this result to \sin can be found in \cite{Horrocks98k}.

Establishing a \pspace-result for \si is not as straightforward as it might seem
at a first glance. One problem is the presence of inverse roles which might lead
to constraints propagating upwards in the tree. This is not compatible with the
standard trace technique \cite{SSSm91} that keeps only a single path in memory
at the same time, because constraints propagating upwards in the tree may have
an influence on paths that have already been visited and have been discarded
from memory. There are at least two possibilities to overcome this problem: (1)
by guessing which constraints might propagate upwards beforehand; (2) by a
\emph{reset-restart} extension of the trace technique described later in this
section.  Unfortunately, this is not the only problem.  To apply either of these
two techniques, it is also necessary to establish a polynomial bound on the
length of paths in the completion tree. This is easily established for logics
such as \alc that do not allow for transitive roles.  For \alc with transitive
roles (i.e., \s), this bound is due to the fact that, for a node $x$ to block a
node $y$, it is sufficient that $\Lab(y)\subseteq \Lab(x)$.
In the presence of inverse roles, we use a more sophisticated blocking technique
to establish the polynomial bound.

\begin{definition}\label{def:si-algo}
  A \emph{completion tree} for an \si concept $D$ is a tree where each node $x$
  of the tree is labelled with two sets $\BLab(x) \subseteq \Lab(x) \subseteq
  \textit{sub}(D)$, and each edge $\tuple{x}{y}$ is labelled with a (possibly
  inverse) role $\Lab(\tuple{x}{y})$ occurring in $\textit{sub}(D)$.

  $R$-neighbours, -successors, and -predecessors are defined as in
  Definition~\ref{def:shin-algo} where, in the absence of role hierarchies,
  $\sss$ is the identity on \Roles.

  A node $x$ is \emph{blocked} iff $x$ has a blocked ancestor $y$, or $x$ has an
  ancestor $y$ and a predecessor $x'$ with $\Lab(\tuple{x'}{x})=S$, and
  $$\BLab(x) \subseteq \Lab(y)\quad  \text{ and }\quad \Lab(x)/\Inv(S) =
  \Lab(y)/\Inv(S),$$
  where $\Lab(x)/\Inv(S) = \{ \all{\Inv(S)}{C} \in \Lab(x)\}$.

  For a node $x$, $\Lab(x)$ is said to contain a \emph{clash} iff $\{ A, \neg A
  \} \subseteq \Lab(x)$. %A completion tree where no node of it contains a clash
  %is called \emph{clash-free}.
  A completion tree to which none of the expansion
  rules given in Figure~\ref{table:alci} is applicable is called \emph{complete}.

  For an \si-concept $D$, the algorithm starts with a
  completion tree consisting of a single node $x$ with
  $\BLab(x) = \Lab(x) =
  \{D\}$. It applies the expansion rules in Figure~\ref{table:alci}, stopping when
  a clash occurs, and answers ``$D$ is satisfiable'' iff the completion rules
  can be applied in such a way that they yield a complete and clash-free
  completion tree.
\end{definition}

\begin{figure}[tb]
  \begin{center}
    \small{
    \framebox[\textwidth]{
    \begin{tabular}{lll}%{|l@{$\;$}l@{$\;$}l|}
      %\hline
      $\sqcap$-rule: & if \hfill 1. & $C_1 \sqcap C_2 \in \Lab(x)$ and \\ &
      \hfill 2. & $\{C_1,C_2\} \not\subseteq \Lab(x)$ \\ & then & $\Lab(x)
      \longrightarrow \Lab(x) \cup \{C_1,C_2\}$ \\

      $\sqcup$-rule: & if \hfill 1. & $C_1 \sqcup C_2 \in \Lab(x)$ and \\ &
      \hfill 2. & $\{C_1,C_2\} \cap \Lab(x) = \emptyset$ \\ & then & $\Lab(x)
      \longrightarrow \Lab(x) \cup \{C\}$ for some $C\in \{C_1,C_2 \}$\\
      $\forall$-rule: & if \hfill 1. & $\all{S}{C} \in \Lab(x)$ and \\ & \hfill
      2. &there is an $S$-successor $y$ of $x$ with $C \notin \BLab(y)$ \\ &
      then &
      $\Lab(y) \longrightarrow \Lab(y) \cup \{C\}$ and \\
      & & $\BLab(y) \longrightarrow \BLab(y) \cup \{C\}$ or\\ & \hfill 2'.&there
      is an $S$-predecessor $y$ of $x$ with $C \notin \Lab(y)$ \\ & then &
      $\Lab(y) \longrightarrow \Lab(y) \cup
      \{C\}$. \\
      $\forall_+$-rule: & if \hfill 1. & $\all{S}{C} \in \Lab(x)$ and $\Tr(S)$
      and \\ & \hfill 2. &there is an $S$-succ.\ $y$ of $x$ with $\all{S}{C}
      \notin \BLab(y)$ \\ & then & $\Lab(y) \longrightarrow \Lab(y) \cup
      \{\all{S}{C}\}$ and \\
      & & $\BLab(y) \longrightarrow \BLab(y) \cup \{\all{S}{C}\}$ or \\ & \hfill
      2'. &there is an $S$-predecessor \ $y$ of $x$ with $\all{S}{C} \notin
      \Lab(y)$ \\ & then & $\Lab(y) \longrightarrow
      \Lab(y) \cup \{\all{S}{C}\}$. \\
      $\exists$-rule: & if \hfill 1. & $\some{S}{C} \in \Lab(x)$, $x$ is not
      blocked and no other rule \\ && is applicable to any of its ancestors, and
      \\ & \hfill 2. & $x$ has no $S$-neighbour $y$ with $C\in \Lab(y)$ \\ &
      then & create a new node $y$ with %\\
      %& &
      $\Lab(\tuple{x}{y})=S$ and $\Lab(y)=
      \BLab(y) = \{C\}$ \\
      %\hline
    \end{tabular}}
    }
    \caption{Tableaux expansion rules for \alci}
    \label{table:alci}\vspace{-0.5cm}
  \end{center}
\end{figure}

As for \shiq, correctness of the algorithm can be proved by first showing that a
\si-concept is satisfiable iff it has a tableau, and next proving the
\si-analogue of Lemma~\ref{lemma:shin-algo-correct}, see \cite{Horrocks98k}.

\begin{theorem}\label{theo:si-algo}
  The tableaux algorithm is a decision procedure for satisfiability and
  subsumption of \si-concepts.
\end{theorem}

Since blocking plays a major r\^ole both in the proof of
Theorem~\ref{theo:si-algo} and especially in the following complexity
considerations, we will discuss it here in more detail.  Blocking guarantees the
termination of the algorithm.  For DLs such as \alc, termination is mainly due
to the fact that the expansion rules can only add new concepts that are strictly
smaller than the concept that triggered their application.

For \s this is no longer true: the $\forall_+$-rule introduces new concepts that
are the same size as the triggering concept. To ensure termination, nodes
labelled with a subset of the label of an ancestor are \emph{blocked}. Since
rules can be applied ``top-down'' (successors are only generated if no other
rules are applicable, and the labels of inner nodes are never touched again) and
subset-blocking is sufficient (i.e., for a node $x$ to be blocked by an ancestor
$y$, it is sufficient that $\Lab(x)\subseteq \Lab(y)$), it is possible to give a
polynomial bound on the length of paths.

For \si, \emph{dynamic blocking} was introduced in~\cite{HoSat98c}, i.e., blocks
are not established on a once-and-for-all basis, but established and broken
dynamically. Moreover, blocks must be established on the basis of label
\emph{equality}, since value restrictions can now constrain predecessors as well as
successors. Unfortunately, this may lead to completion trees with
exponentially long paths because there are exponentially many possibilities to
label sets on such a path.  Due to the non-deterministic $\sqcup$-rule, these
exponentially many sets may actually occur.

This non-determinism is not problematical for $\mathcal{S}$ because disjunctions
need not be completely decomposed to yield a subset-blocking situation.  For an
optimal \si algorithm, the additional label $\BLab$ was introduced to enable a
sort of subset-blocking which is independent of the $\sqcup$-non-determinism.
Intuitively, $\BLab(x)$ is the restriction of $\Lab(x)$ to those non-decomposed
concepts that $x$ must satisfy, whereas $\Lab(x)$ contains boolean
decompositions of these concepts as well as those that are imposed by value
restrictions in descendants.  If $x$ is blocked by $y$, then all concepts in
$\BLab(x)$ are eventually decomposed in $\Lab(y)$. However,
in order to substitute $x$ by $y$, $x$'s constraints on predecessors must be at
least as strong as $y$'s; this is taken care of by the second blocking
condition.

Let us consider a path $x_0, x_1,\dots,x_n$ where all edges are labelled $R$
with $\Tr(R)$, the only kind of path along which the length of the longest
concept in the labels might not decrease. If no rules can be applied, then we
have, for $1 \leq i < n$,
$$
\begin{array}[t]{rcl}
  \Lab(x_{i+1})/\Inv(R) &  \subseteq& \Lab(x_{i})/\Inv(R) \mbox{\ \ and}\\
  \BLab(x_i)&\subseteq &\BLab(x_{i+1}) \cup\{C_i\}
\end{array}$$
(where $\some{R}{C_i} \in \Lab(x_i)$ triggered the generation of $x_{i+1}$).
This limits the number of different labels and guarantees blocking after a
polynomial number of steps.

\begin{lemma}\label{lemma:poly_paths}
The paths of a completion tree for a concept $D$ have a length of at most $m^4$
where $m = |\textit{sub}(D)|$.
\end{lemma}

Finally, a slight modification of the expansion rules given in
Figure~\ref{table:alci} yields a \pspace algorithm.
This modification is necessary because the original algorithm must keep the
whole completion tree in memory---which needs exponential space even though the
length of its paths is polynomially bounded. The original algorithm may not
forget about branches because restrictions which are pushed \emph{upwards} in
the tree might make it necessary to revisit paths which have been considered
before. A \emph{reset-restart} mechanism solves this problem as follows:

Whenever the $\forall$- or the $\forall_+$-rule is applied to a node $x$ and its
\emph{predecessor} $y$ (Case 2' of these rules), we delete all successors of $y$
from the completion tree (\emph{reset}). While this makes it necessary to
\emph{restart} the generation of successors for $y$, it makes it possible to
implement the algorithm in a depth-first manner which facilitates the re-use of
space.

This modification does not affect the proof of soundness and completeness for
the algorithm, but of course we have to re-prove termination \cite{Horrocks98k}
as it formerly relied on the fact that we never removed any nodes from the
completion tree.  Summing up we get:
\begin{theorem}
  The modified algorithm is a \pspace decision procedure for satisfiability
  and subsumption of \si-concepts.
\end{theorem}

\newcommand{\shinplus}{\ensuremath{\mathcal{SHIN}^+}\xspace}

\section{The Undecidability of Unrestricted \shin}

Like earlier DLs that combine a hierarchy of (transitive and non-transitive)
roles with some form of number restrictions~\cite{HoSat98c,Horrocks98k}, \shin
only allows \emph{simple} roles in restrictions, i.e. roles that are neither
transitive nor have transitive subroles. The justification for this limitation
has been partly on the grounds of a doubtful semantics (of transitive functional roles) and
partly to simplify decision procedures. In this section, we will show that
allowing arbitrary roles in \shin number restrictions leads to undecidability.
For convenience, we denote \shin with arbitrary roles in number restrictions by
\shinplus.

The undecidability proof uses a reduction of the domino problem~\cite{Ber66}
adapted from~\cite{Baader96b}. This problem asks whether, for a set of domino types,
there exists a \emph{tiling} of an $\N^2$ grid such that each point of the grid
is covered with exactly one of the domino types, and adjacent dominoes are
``compatible'' with respect to some predefined criteria.

\begin{definition}
  A domino system $\mathcal{D} = (D, H, V)$ consists of a non-empty
  set of domino types $D = \{D_1,\ldots,D_n\}$, and of sets of
  horizontally and vertically matching pairs $H \subseteq D \times D$
  and $V \subseteq D \times D$. The problem is to determine if, for a
  given $\mathcal{D}$, there exists a \emph{tiling} of an $\N \times
  \N$ grid such that each point of the grid is covered with a domino
  type in $D$ and all horizontally and vertically adjacent pairs of
  domino types are in $H$ and $V$ respectively, i.e., a mapping $t:\N
  \times \N \rightarrow D$ such that for all $m,n \in \N$,
  $\tuple{t(m,n)}{t(m+1,n)} \in H$ and $\tuple{t(m,n)}{t(m,n+1)} \in
  V$.
\end{definition}

This problem can be reduced to the satisfiability of \shinplus-concepts, and the
undecidability of the domino problem implies undecidability of satisfiability of
\shinplus-concepts.

Ensuring that each point is associated with exactly one domino type and that a
point and its neighbours satisfy the compatibility conditions induced by $H$ and
$V$ is simple for most logics (via the introduction of concepts $C_{D_i}$ for domino types
$D_i$, and the use of value restrictions and boolean connectives), and applying such
conditions throughout the grid is also simple in a logic such as \shinplus which
can deal with arbitrary axioms.  The crucial difficulty is representing the $\N
\times \N$ grid using ``horizontal'' and ``vertical'' roles $X$ and $Y$, and in
particular forcing the coincidence of $X \circ Y$- and $Y \circ X$-successors.
This can be accomplished in \shinplus using an alternating pattern of two
horizontal roles $X_1$ and $X_2$, and two vertical roles $Y_1$ and $Y_2$, with
disjoint primitive concepts $A$, $B$, $C$, and $D$ being used to identify points
in the grid with different combinations of successors. The coincidence of $X
\circ Y$ and $Y \circ X$ successors can then be enforced using number
restrictions on transitive super-roles of each of the four possible combinations
of $X$ and $Y$ roles. A visualisation of the resulting grid and a suitable role
hierarchy is shown in Figure~\ref{fig:grid}, where $S^\oplus_{ij}$ are
transitive roles.

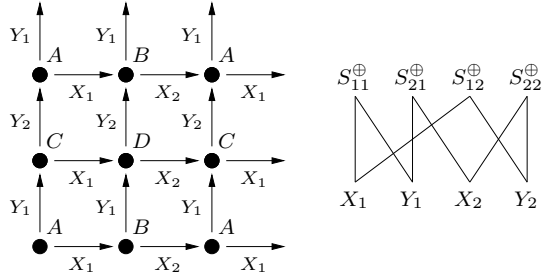
\begin{figure}%[t!]
  \begin{center}
    \parbox[t]{3.2in}{\input{tiling.pstex_t}}
    \caption{Visualisation of the grid and role hierarchy.}
    \label{fig:grid}\vspace{-0.5cm}
  \end{center}
\end{figure}

The alternation of $X$ and $Y$ roles in the grid means that one of the
transitive super-roles $S_{ij}$ connects each point $(m, n)$ to the points
$(m+1, n)$, $(m, n+1)$ and $(m+1, n+1)$, and to no other points. A number
restriction of the form \atmost{3}{S_{ij}} can thus be used to enforce the
necessary coincidence of $X \circ Y$- and $Y \circ X$-successors.  A complete
specification of the grid is given by the following axioms:
{\setlength{\arraycolsep}{0em}$$\begin{array}{rcl}
A & \mbox{}\sqsubseteq\mbox{} & \neg B \sqcap \neg C \sqcap \neg
  D \sqcap \some{X_1}{B} \sqcap \some{Y_1}{C}
  \sqcap \atmost{3}{S_{11}}\mbox{,} \\
B & \sqsubseteq & \neg A \sqcap \neg C \sqcap \neg
  D \sqcap \some{X_2}{A} \sqcap \some{Y_1}{D}
  \sqcap \atmost{3}{S_{21}}\mbox{,} \\
C & \sqsubseteq & \neg A \sqcap \neg B \sqcap \neg
  D \sqcap \some{X_1}{D} \sqcap \some{Y_2}{A}
  \sqcap \atmost{3}{S_{12}}\mbox{,} \\
D & \sqsubseteq & \neg A \sqcap \neg B \sqcap \neg
  C \sqcap \some{X_2}{C} \sqcap \some{Y_2}{B}
  \sqcap \atmost{3}{S_{22}}\mbox{.}
\end{array}$$}
It only remains to add axioms which encode the local compatibility conditions
(as described in ~\cite{Baader96b}) and to assert that $A$, $B$, $C$, and $D$
are subsumed by the disjunction of all domino types to enforce the placement of
a tile on each point of the grid. The concept $A$ is now satisfiable w.r.t.\ the
various axioms (which can be internalised as described in
Lemma~\ref{lemma:terminologies}) iff there is a compatible tiling of the grid.

\newcommand{\IFact}{I-FaCT\xspace}

\section{Discussion}

A new DL system is being implemented based
on the \shiq algorithm described in Section~\ref{sec:shin-algo}. Pending the
completion of this project, the existing \Fact system~\cite{Horrocks98c} has
been modified to deal with inverse roles using the \shiq
blocking strategy, giving a DL which is equivalent to \shi extended with
functional roles~\cite{HoSat98c}; we will refer to this DL as \shif and to the
modified \Fact system as \IFact.

\IFact has been used to conduct some initial experiments with a terminology
representing (fragments of) database schemata and inter schema assertions from a
data warehousing application~\cite{Calvanese98c}
(a slightly simplified version of the proposed encoding was used to generate
\shif terminologies).
\IFact is able to classify this terminology,
which contains 19 concepts and 42 axioms, in less than 0.1s of (266MHz Pentium)
CPU time. In contrast, eliminating inverse roles using an embedding
technique~\cite{CaGiaRo98} gives an equisatisfiable \Fact terminology with an additional
84 axioms, but one which \Fact is unable to classify in 12 hours of CPU time.

An extension of the embedding technique can be used to eliminate number
restrictions~\cite{DeGiacomo95b}, but requires a target logic which
supports the transitive \emph{closure} of roles, i.e., \CPDL. The even larger
number of axioms which this embedding would introduce makes it unlikely that
tractable reasoning could be performed on the resulting terminology. Moreover,
we are not aware of any algorithm for \CPDL which does not employ a so-called
\emph{cut rule}~\cite{DeGiaMass98}, the application of which introduces
considerable additional non-determinism. It seems inevitable that this would
lead to a further degradation in empirical tractability.

As far as complexity is concerned, we have already been successful in extending the \pspace-result for \si to
\sin~\cite{Horrocks98k}. Currently we are working on an extension of this result
to \siq combining the techniques from this paper with those presented in
\cite{Tobies-CADE-99}.

\newcommand{\etalchar}[1]{$^{#1}$}

\clearpage
\section*{Appendix}

In this appendix we present the proof of
Lemma~\ref{lemma:shin-algo-correct}, which is repeated here for easier
reference.

\vspace{1ex}

\noindent \textbf{Lemma.} {\em Let $D$ be an \shiq-concept.
  \begin{enumerate}
  \item (Termination) The tableaux algorithm terminates when started with $D$.
  \item (Soundness) If the expansion rules can be applied to $D$ such that they
    yield a complete and clash-free completion tree, then $D$ has a tableau.
  \item (Completeness) If $D$ has a tableau, then the expansion rules can be
    applied to $D$ such that they yield a complete and clash-free completion
    tree.
  \end{enumerate}
}

\paragraph{(Termination)}
Let $m = \Card{\clos(D)}$, $k = \Card{\Roles_D}$, and $n_{\textit{max}}$ the
maximum $n$ that occurs in a concept of the form $(\bowtie \; n \; S \; C) \in
\clos(D)$.  Termination is a consequence of the following properties of the
expansion rules:

\begin{itemize}
\item The expansion rules never remove nodes from the tree or concepts from node
  labels. Edge labels can only be changed by the $\leqslant$-rule which either
  expands them or sets them to $\emptyset$; in the latter case the node below
  the $\emptyset$-labelled edge is blocked and this block is never broken.
\item Each successor of a node $x$ is the result of the application of the
  $\exists$-rule or the $\geqslant$-rule to $x$. For a node $x$, each concept in
  $\Lab(x)$ can trigger the generation of successors at most once.

  For the $\exists$-rule, if a successor $y$ of $x$ was generated for a concept
  $\some S C \in \Lab(x)$ and later $\Lab(\tuple x y)$ is set to $\emptyset$ by
  the $\leqslant$-rule, then there is some $S$-neighbour $z$
  of $x$ with $C \in \Lab(z)$.

  For the $\geqslant$-rule, if $y_1, \dots, y_n$ were generated by the
  $\geqslant$-rule for $\qnrgeq n S C\in\Lab(x)$, then $y_i \ndoteq y_j$ holds
  for all $1 \leq i < j \leq n$.  This implies that there are always $n$
  $S$-neighbours $y'_1, \dots, y'_n$ of $x$ with $C \in \Lab(y'_i)$ and $y'_i
  \ndoteq y'_j$ for all $1 \leq i < j \leq n$, since the $\leqslant$-rule never
  merges two nodes $y'_i, y'_j$ with $y'_i \ndoteq y'_j$, and, whenever an
  application of the $\leqslant$-rule sets $\Lab(\tuple x {y'_i})$ to
  $\emptyset$, there is some $S$-neighbour $z$ of $x$ which ``inherits'' both
  $C$ and all inequalities from $y'_i$.

  Since $\clos(D)$ contains a total of at most $m$ \some{R}{C} and $\qnrgeq n
  S C$ concepts, the out-degree of the tree is bounded by $m\cdot
  n_{\textit{max}}$.

\item Nodes are labelled with non-empty subsets of $\clos(D)$ and edges with
  subsets of $R_D$, so there are at most $2^{2mk}$ different possible labellings
  for a pair of nodes and an edge.  Therefore, if a path $p$ is of length at
  least $2^{2mk}$, then from the pair-wise blocking condition there must be two
  nodes $x,y$ on $p$ such that $x$ is directly blocked by $y$.  Furthermore, if
  a node was generated at distance $\ell$ from the root node, it always remains
  at this distance, and thus paths are not curled up or shortened.
  Since a path on which nodes are blocked cannot become longer, paths are of
  length at most $2^{2mn}$. \qed
  \end{itemize}

\paragraph{(Soundness)}
Let \Tree be a complete and clash-free completion tree. A path is a sequence
of pairs of nodes of \Tree of the form $p = [\frac{x_0}{x'_0}, \dots,
\frac{x_n}{x'_n}]$. For such a path we define $\Tail(p) := x_n$ and $\Tail'(p)
:= x'_n$. With $[p|\frac{x_{n+1}}{x'_{n+1}}]$ we denote the path
$[\frac{x_0}{x'_0}, \dots, \frac{x_n}{x'_n}, \frac{x_{n+1}}{x'_{n+1}}]$. The
set $\Paths(\Tree)$ is defined inductively as follows:
  \begin{itemize}
  \item For the root node $x_0$ of \Tree, $[\frac{x_0}{x_0}] \in
    \Paths(\Tree)$, and
  \item For a path $p \in \Paths(\Tree)$ and a node $z$ in \Tree:
    \begin{itemize}
    \item if $z$ is a successor of $\Tail(p)$ and $z$ is not blocked, then
      $[p|\frac{z}{z}] \in \Paths(\Tree)$, or
    \item if, for some node $y$ in \Tree, $y$ is a successor of $\Tail(p)$ and
      $z$ blocks $y$, then $[p|\frac{z}{y}] \in \Paths(\Tree)$.
    \end{itemize}
  \end{itemize}

  Please note that, due to the construction of \Paths, for $p \in \Paths(\Tree)$
  with $p=[p'|\frac{x}{x'}]$, we have that $x$ is not blocked, $x'$ is blocked
  iff $x \neq x'$, and $x'$ is never indirectly blocked. Furthermore, $\Lab(x) =
  \Lab(x')$ holds.

  Now we can define a tableau $T = (\mathbf{S},\Lab,\Edges)$ with:
  \[
  \hspace{-1cm}\begin{array}{r@{\,}c@{\,}l}
    \mathbf{S} & = & \Paths(\Tree)\\[0.5ex]
    \Lab(p) & = & \Lab(\Tail(p))\\[0.5ex]
    \Edges(R) & = & \{\tuple{p}{q} \in \mathbf{S} \times \mathbf{S} \mid
    \begin{array}[t]{l@{\,}l}
      \multicolumn{2}{l}{\mbox{Either $q = [p|\frac{x}{x'}]$ and}}\\
      \quad \text{$x'$ is an $R$-successor of $\Tail(p)$}\\
      \multicolumn{2}{l}{\mbox{or $p = [q|\frac{x}{x'}]$  and}}\\
      \quad \text{$x'$ is an $\Inv(R)$-successor of $\Tail(q)$}\}.
    \end{array}
  \end{array}
  \]
  \noindent \textsc{Claim:} $T$ is a tableau for $D$ with respect to ${\mathcal{R}^+}$.
  \vspace{1ex}

  \noindent We show that $T$ satisfies all the properties from
  Definition~\ref{def:alchr2}.

  \begin{itemize}
  \item $D \in \Lab([\frac{x_0}{x_0}])$ since $D \in \Lab(x_0)$.
  \item \textbf{Property 1} holds because \Tree is clash-free;
    \textbf{Properties 2,3} hold because $\Tail(p)$ is not blocked and \Tree
    is complete.
  \item \textbf{Property 4}: Assume $\forall S.C \in \Lab(p)$ and $\tuple p q
    \in \Edges(S)$. If $q = [p | \frac{x}{x'}]$, then $x'$ is an $S$-successor
    of $\Tail(p)$ and thus $C \in \Lab(x')$ (because the $\forall$-rule is not
    applicable). Since $\Lab(q) = \Lab(x) = \Lab(x')$, we have $C \in \Lab(q)$.
    If $p = [q | \frac{x}{x'}]$, then $x'$ is an $\Inv(S)$-successor of
    $\Tail(q)$ and thus $C \in \Lab(\Tail(q))$ (because $x'$ is not indirectly
    blocked and the $\forall$-rule is not applicable), hence $C \in \Lab(q)$.
  \item \textbf{Property 5}: Assume $\some S C \in \Lab(p)$. Define $x :=
    \Tail(p)$.  In \Tree there is an $S$-neighbour $y$ of $x$ with $C \in
    \Lab(y)$, because the $\exists$-rule is not applicable. There are two
    possibilities:
    \begin{itemize}
    \item $y$ is a successor of $x$ in \Tree. If $y$ is not blocked, then $q
      := [p|\frac{y}{y}] \in \mathbf{S}$ and $\tuple p q \in \Edges(S)$ as
      well as $C \in \Lab(q)$. If $y$ is blocked by some node $z$ in \Tree,
      then $q := [p|\frac{z}{y}] \in \mathbf{S}$.
    \item $y$ is a predecessor of $x$. Again, there are two possibilities:
      \begin{itemize}
      \item $p$ is of the form $p = [q|\frac{x}{x'}]$ with $\Tail(q) = y$.
      \item $p$ is of the form $p = [q|\frac{x}{x'}]$ with $\Tail(q) = u \neq
        y$. $x$ only has one predecessor in $\Tree$, hence $u$ is not the
        predecessor of $x$. This implies $x \neq x'$, $x$ blocks $x'$ in \Tree,
        and $u$ is the predecessor of $x'$ due to the construction of $\Paths$.
        Together with the definition of the blocking condition, this implies
        $\Lab(\tuple u {x'}) = \Lab(\tuple y x)$ as well as $\Lab(u) = \Lab(y)$
        due to the pair-wise blocking condition.
      \end{itemize}
    \end{itemize}
    In all three cases, $\tuple p q \in \Edges(S)$ and $C \in \Lab(q)$.
  \item \textbf{Property 6}: Assume $\forall S.C \in \Lab(p)$, $\tuple p q \in
    \Edges(R)$ for some $R \sss S$ with $\Tr(R)$. If $q = [p|\frac{x}{x'}]$,
    then $x'$ is an $R$-successor of $\Tail(p)$ and thus $\forall R.C \in
    \Lab(x')$ (because otherwise the $\forall_+$-rule would be applicable). From
    $\Lab(q) = \Lab(x) = \Lab(x')$ it follows that $\forall R.C \in \Lab(q)$. If
    $p = [q|\frac{x}{x'}]$, then $x'$ is an $\Inv(S)$-successor of $\Tail(q)$
    and hence $\Tail(q)$ is an $R$-neighbour of $x'$. Because $x'$ is not
    indirectly blocked, this implies $\forall R.C \in \Lab(\Tail(q))$ and hence
    $\forall R.C \in \Lab(q)$.
  \item \textbf{Property 11}: Assume $(\bowtie \; n \; S \; C) \in \Lab(p)$,
    $\tuple p q \in \Edges(S)$. If $q = [p | \frac{x}{x'}]$, then $x'$ is an
    $S$-successor of $\Tail(p)$ and thus $\{C,\nneg C\} \cap \Lab(x') \neq
    \emptyset$ (since the \textit{choose}-rule is not applicable).  Since
    $\Lab(q) = \Lab(x) = \Lab(x')$, we have $\{C, \nneg C\} \cap \Lab(q) \neq
    \emptyset$.  If $p = [q | \frac{x}{x'}]$, then $x'$ is an
    $\Inv(S)$-successor of $\Tail(q)$ and thus $\{C, \nneg C\} \cap
    \Lab(\Tail(q)) \neq \emptyset$  (since $x'$ is not indirectly
    blocked and the \textit{choose}-rule is not applicable), hence $\{C, \nneg
    C\} \cap \Lab(q) \neq \emptyset$.
  \item Assume \textbf{Property 9} is violated. Hence there is some $p \in
    \mathbf{S}$ with $\qnrleq n S C \in \Lab(p)$ and $\sharp S^T(p,C) > n$. We
    show that this implies $\sharp S^\Tree(\Tail(p),C) > n$, in
    contradiction of either the clash-freeness or completeness of \Tree.
    Define $x := \Tail(p)$ and $P := S^T(p,C)$.  Due to the assumption, we
    have $\sharp P > n$.  We distinguish two cases:
    \begin{itemize}
    \item $P$ contains only paths of the form $q = [p|\frac{y}{y'}]$. We claim
      that the function $\Tail'$ is injective on $P$. Assume that there are two
      paths $q_1,q_1 \in P$ with $q_1 \neq q_2$ and $\Tail'(q_1) = \Tail'(q_2) =
      y'$. Then $q_1$ is of the form $q_1 = [p | (y_1,y')]$ and $q_2$ is of the
      form $q_2 = [p | \frac{y_2}{y'}]$ with $y_1 \neq y_2$. If $y'$ is not
      blocked in \Tree, then $y_1 = y' = y_2$, contradicting $y_1 \neq
      y_2$.  If $y'$ is blocked in \Tree, then both $y_1$ and $y_2$ block $y'$,
      which implies $y_1 = y_2$, again a contradiction.

      Since $\Tail'$ is injective on $P$, it holds that $\sharp P = \sharp
      \Tail'(P)$. Also for each $y' \in \Tail'(P)$, $y'$ is an $S$-successor
      of $x$ and $C \in \Lab(y')$. This implies $\sharp S^\Tree(x,C) > n$.
    \item $P$ contains a path $q$ where $p$ is of the form $p = [q |
      \frac{x}{x'}]$.  Obviously, $P$ may only contain one such path. As in
      the previous case, $\Tail'$ is an injective function on the set $P' := P
      \setminus \{ q \}$, each $y' \in \Tail'(P')$ is an $S$-successor of $x$
      and $C \in \Lab(y')$ for each $y' \in \Tail'(P')$. To show that indeed
      $\sharp S^\Tree(x,C)>n$ holds, we have to prove the existence of a
      further $S$-neighbour $u$ of $x$ with $C \in \Lab(u)$ and $u \not\in
      \Tail'(P')$.  This will be ``supplied'' by $z:= \Tail(q)$.  We
      distinguish two cases:
      \begin{itemize}
      \item $x=x'$. Hence $x$ is not blocked. This implies that $x$ is an
        $\Inv(S)$-successor of $z$ in \Tree. Since $\Tail'(P')$ contains only
        successors of $x$, we have that $z \not\in \Tail'(P')$ and, by
        construction, $z$ is an $S$-neighbour of $x$ with $C \in \Lab(z)$.
      \item $x \neq x'$. This implies that $x'$ is blocked in \Tree by $x$ and
        that $x'$ is an $\Inv(S)$-successor of $z$ in \Tree. The definition of
        pairwise-blocking implies that $x$ is an $\Inv(S)$-successor of some
        node $u$ in \Tree with $\Lab(u) = \Lab(z)$. Again, since $\Tail'(P')$
        contains only successors of $x$ we have that $u \not\in \Tail'(P')$ and,
        by construction, $u$ is an $S$-neighbour of $x$ and $C \in \Lab(u)$.
      \end{itemize}
    \end{itemize}
  \item \textbf{Property 10}: Assume $\qnrgeq n S C \in \Lab(p)$. Completeness
    of \Tree implies that there exist $n$ individuals $y_1, \dots, y_n$ in \Tree
    such that each $y_i$ is an $S$-neighbour of $\Tail(p)$ and $C \in
    \Lab(y_i)$. We claim that, for each of these individuals, there is a path
    $q_i$ such that $\tuple p {q_i} \in \Edges(S)$, $C \in \Lab(q_i)$, and $q_i
    \neq q_j$ for all $1 \leq i < j \leq n$. Obviously, this implies $\sharp
    S^T(p,C) \geqslant n$.  For each $y_i$ there are three possibilities:
    \begin{itemize}
    \item $y_i$ is an $S$-successor of $x$ and $y_i$ is not blocked in \Tree.
      Then $q_i = [p|\frac{y_i}{y_i}]$ is a path with the desired properties.
    \item $y_i$ is an $S$-successor of $x$ and $y_i$ is blocked in \Tree by
      some node $z$. Then $q_i = [p|\frac{z}{y_i}]$ is the path with the
      desired properties. Since the same $z$ may block several of the $y_j$s,
      it is indeed necessary to include $y_i$ explicitly into the path to make
      them distinct.
    \item $x$ is an $\Inv(S)$-successor of $y_i$. There may be at most one
      such $y_i$. This implies that $p$ is of the form $p = [q |
      \frac{x}{x'}]$ with $\Tail(q) = y_i$.  Again, $q$ has the desired
      properties and, obviously, $q$ is distinct from all other paths $q_j$.
    \end{itemize}
  \item \textbf{Property 7} is satisfied due to the symmetric definition of
    $\Edges$. \textbf{Property 8} is satisfied due to the definition of
    $R$-successor that takes into account the role hierarchy $\sss$. \qed
  \end{itemize}

\paragraph{(Completeness)}
Let $T = (\mathbf{S},\Lab,\Edges)$ be a tableau for $D$ w.r.t.
${\mathcal{R}^+}$. We use this tableau to guide the application of the
non-deterministic rules. To do this, we will inductively define a function
$\pi$, mapping the individuals of the tree \Tree to $\mathbf{S}$ such that, for
each $x,y$ in \Tree:
  \[
  \left.
    \begin{array}{l}
      \Lab(x) \subseteq \Lab(\pi(x))\\
      \text{if $y$ is an $S$-neighbour of $x$, then $\tuple {\pi(x)} {\pi(y)} \in
        \Edges(S)$}\\
      \text{$x \ndoteq y$ implies $\pi(x) \neq \pi(y)$}
    \end{array}
    \qquad \right \} (*)
  \]

  \noindent \textsc{Claim:} Let \Tree be a completion-tree and $\pi$ a function that
  satisfies $(*)$. If a rule is applicable to $\Tree$ then the rule is
  applicable to $\Tree$ in a way that yields a completion-tree $\Tree'$ and an
  extension of $\pi$ that satisfy $(*)$.  \vspace{1ex}

  \noindent Let \Tree be a completion-tree and $\pi$ be a function that satisfies $(*)$.
  We have to consider the various rules.
  \begin{itemize}
  \item \textbf{The $\sqcap$-rule:} If $C_1 \sqcap C_2 \in \Lab(x)$, then $C_1
    \sqcap C_2 \in \Lab(\pi(x))$. This implies $C_1, C_2 \in \Lab(\pi(x))$ due
    to Property 2 from Definition~\ref{def:alchr2}, and hence the rule can be
    applied without violating $(*)$.
  \item \textbf{The $\sqcup$-rule:} If $C_1 \sqcup C_2 \in \Lab(x)$, then $C_1
    \sqcup C_2 \in \Lab(\pi(x))$. Since $T$ is a tableau, Property 3 from
    Definition \ref{def:alchr2} implies $\{ C_1, C_2 \} \cap \Lab(\pi(x)) \neq
    \emptyset$. Hence the $\sqcup$-rule can add a concept $E \in \{ C_1, C_2
    \}$ to $\Lab(x)$ such that $\Lab(x) \subseteq \Lab(\pi(x))$ holds.
  \item \textbf{The $\exists$-rule:} If $\some S C \in \Lab(x)$, then $\some S C
    \in \Lab(\pi(x))$ and, since $T$ is a tableau, Property 5 of Definition
    \ref{def:alchr2} implies that there is an element $t \in \mathbf{S}$ such
    that $\tuple {\pi(x)} t \in \Edges(S)$ and $C \in \Lab(t)$.  The application
    of the $\exists$-rule generates a new variable $y$ with $\Lab(\tuple x y =
    \{ S \}$ and $\Lab(y) = \{C\}$. Hence we set $\pi := \pi[y \mapsto t]$ which
    yields a function that satisfies $(*)$ for the modified tree.
  \item \textbf{The $\forall$-rule:} If $\forall S.C \in \Lab(x)$, then
    $\forall S.C \in \Lab(\pi(x))$, and if $y$ is an $S$-neighbour of $x$,
    then also $\tuple{\pi(x)}{\pi(y)} \in \Edges(S)$ due to $(*)$. Since $T$
    is a tableau, Property 4 of Definition~\ref{def:alchr2} implies $C \in
    \Lab(\pi(y))$ and hence the $\forall$-rule can be applied without
    violating $(*)$.
  \item \textbf{The $\forall_+$-rule:} If $\forall S.C \in \Lab(x)$, then
    $\forall S.C \in \Lab(\pi(x))$, and if there is some $R \sss S$ with
    $\Tr(R)$ and $y$ is an $R$-neighbour of $x$, then also
    $\tuple{\pi(x)}{\pi(y)} \in \Edges(R)$ due to $(*)$. Since $T$ is a
    tableau, Property 6 of Definition \ref{def:alchr2} implies $\forall R.C
    \in \Lab(\pi(y))$ and hence the $\forall_+$-rule can be applied without
    violating $(*)$.
  \item \textbf{The \textit{choose}-rule:} If $(\bowtie \; n \; S \; C) \in
    \Lab(x)$, then $(\bowtie \; n \; S \; C) \in \Lab(\pi(x))$, and, if there
    is an $S$-neighbour $y$ of $x$, then $\tuple {\pi(x)} {\pi(y)} \in
    \Edges(S)$ due to $(*)$. Since $T$ is a tableau, Property 11 of
    Definition~\ref{def:alchr2} implies $\{ C, \nneg C \} \cap \Lab(\pi(y)
    \neq \emptyset$. Hence the \textit{choose}-rule can add an appropriate
    concept $E \in \{ C, \nneg C \}$ to $\Lab(x)$ such that $\Lab(y) \subseteq
    \Lab(\pi(y))$ holds.
  \item \textbf{The $\geqslant$-rule:} If $\qnrgeq n S C \in \Lab(x)$, then
    $\qnrgeq n S C \in \Lab(\pi(x))$. Since $T$ is a tableau, Property 10 of
    Definition \ref{def:alchr2} implies $\sharp S^T(\pi(x),C) \geqslant n$.
    Hence there are individuals $t_1,\dots,t_n \in \mathbf{S}$ such that
    $\tuple {\pi(x)} {t_i} \in \Edges(S)$, $C \in \Lab(t_i)$, and $t_i \neq
    t_j$ for $1 \leq i < j \leq n$. The $\geqslant$-rule generates $n$ new
    nodes $y_1, \dots, y_n$. By setting $\pi := \pi [y_1 \mapsto t_1, \dotsm
    y_n \mapsto t_n]$, one obtains a function $\pi$ that satisfies $(*)$ for
    the modified tree.
  \item \textbf{The $\leqslant$-rule:} If $\qnrleq n S C \in \Lab(x)$, then
    $\qnrleq n S C \in \Lab(\pi(x))$. Since $T$ is a tableau, Property 9 of
    Definition~\ref{def:alchr2} implies $\sharp S^T(\pi(x),C) \leqslant n$. If
    the $\leqslant$-rule is applicable, we have $\sharp S^\Tree(x,C) > n$,
    which implies that there are at least $n+1$ $S$-neighbours $y_0,\dots,y_n$
    of $x$ such that $C \in \Lab(y_i)$. Thus, there must be two nodes $y,z \in
    \{ y_0, \dots, y_n \}$ such that $\pi(y) = \pi(z)$ (because otherwise
    $\sharp S^T(\pi(x),C) > n$ would hold). From $\pi(y) = \pi(z)$ we have that $y
    \ndoteq z$ cannot hold because of $(*)$, and $y,z$ can be chosen such that
    $y$ is not an ancestor of $z$. Hence the $\leqslant$-rule can be applied
    without violating $(*)$.
  \end{itemize}

  Why does this claim yield the completeness of the tableaux algorithm? For the
  initial completion-tree consisting of a single node $x_0$ with $\Lab(x_0) =
  \{D \}$ and ${\ndoteq} = {\emptyset}$ we can give a function $\pi$ that
  satisfies $(*)$ by setting $\pi(x_0) := s_0$ for some $s_0 \in \mathbf{S}$
  with $D \in \Lab(s_0)$ (such an $s_0$ exists since $T$ is a tableau for $D$).
  Whenever a rule is applicable to \Tree, it can be applied in a way that
  maintains $(*)$, and, since the algorithm terminates, we have that any
  sequence of rule applications must terminate. Properties $(*)$ imply that any
  tree \Tree generated by these rule-applications must be clash-free
  as there are only two possibilities for a clash, and it is easy to
  see that neither of these can hold in \Tree:
  \begin{itemize}
  \item \Tree cannot contain a node $x$ such that $\{ C , \neg C \} \in \Lab(x)$
    because $\Lab(x) \subseteq \Lab(\pi(x))$ and hence Property 1 of Definition
    \ref{def:alchr2} would be violated for $\pi(x)$.
  \item \Tree cannot contain a node $x$ with $\qnrleq n S C \in \Lab(x)$ and
    $n+1$ $S$-neighbours $y_0, \dots y_n$ of $x$ with $C \in \Lab(y_i)$ and $y_i
    \ndoteq y_j$ for $0 \leq i < j \leq n$ because $\qnrleq n S C \in
    \Lab(\pi(x))$, and, since $y_i \ndoteq y_j$ implies $\pi(y_i) \neq
    \pi(y_j)$, $\sharp S^T(\pi(x),C) > n$, in contradiction to Property 9 of
    Definition~\ref{def:alchr2}. \qed
  \end{itemize}

\end{document}

%% file: tiling.pstex_t
\begin{picture}(0,0)%
\epsfig{file=tiling.pstex}%
\end{picture}%
\setlength{\unitlength}{2368sp}%
\begingroup\makeatletter\ifx\SetFigFont\undefined%
\gdef\SetFigFont#1#2#3#4#5{%
  \reset@font\fontsize{#1}{#2pt}%
  \fontfamily{#3}\fontseries{#4}\fontshape{#5}%
  \selectfont}%
\fi\endgroup%
\begin{picture}(6375,2787)(649,-6586)
\put(1651,-4786){\makebox(0,0)[b]{\smash{\SetFigFont{7}{8.4}{\rmdefault}{\mddefault}{\itdefault}$X_1$}}}
\put(1276,-4411){\makebox(0,0)[lb]{\smash{\SetFigFont{8}{9.6}{\rmdefault}{\mddefault}{\itdefault}$A$}}}
\put(2026,-4186){\makebox(0,0)[rb]{\smash{\SetFigFont{7}{8.4}{\rmdefault}{\mddefault}{\itdefault}$Y_1$}}}
\put(2926,-4186){\makebox(0,0)[rb]{\smash{\SetFigFont{7}{8.4}{\rmdefault}{\mddefault}{\itdefault}$Y_1$}}}
\put(3451,-4786){\makebox(0,0)[b]{\smash{\SetFigFont{7}{8.4}{\rmdefault}{\mddefault}{\itdefault}$X_1$}}}
\put(3076,-4411){\makebox(0,0)[lb]{\smash{\SetFigFont{8}{9.6}{\rmdefault}{\mddefault}{\itdefault}$A$}}}
\put(2551,-4786){\makebox(0,0)[b]{\smash{\SetFigFont{7}{8.4}{\rmdefault}{\mddefault}{\itdefault}$X_2$}}}
\put(2176,-4411){\makebox(0,0)[lb]{\smash{\SetFigFont{8}{9.6}{\rmdefault}{\mddefault}{\itdefault}$B$}}}
\put(1651,-5686){\makebox(0,0)[b]{\smash{\SetFigFont{7}{8.4}{\rmdefault}{\mddefault}{\itdefault}$X_1$}}}
\put(3451,-5686){\makebox(0,0)[b]{\smash{\SetFigFont{7}{8.4}{\rmdefault}{\mddefault}{\itdefault}$X_1$}}}
\put(2551,-5686){\makebox(0,0)[b]{\smash{\SetFigFont{7}{8.4}{\rmdefault}{\mddefault}{\itdefault}$X_2$}}}
\put(1126,-5986){\makebox(0,0)[rb]{\smash{\SetFigFont{7}{8.4}{\rmdefault}{\mddefault}{\itdefault}$Y_1$}}}
\put(1651,-6586){\makebox(0,0)[b]{\smash{\SetFigFont{7}{8.4}{\rmdefault}{\mddefault}{\itdefault}$X_1$}}}
\put(1276,-6211){\makebox(0,0)[lb]{\smash{\SetFigFont{8}{9.6}{\rmdefault}{\mddefault}{\itdefault}$A$}}}
\put(2026,-5986){\makebox(0,0)[rb]{\smash{\SetFigFont{7}{8.4}{\rmdefault}{\mddefault}{\itdefault}$Y_1$}}}
\put(2926,-5986){\makebox(0,0)[rb]{\smash{\SetFigFont{7}{8.4}{\rmdefault}{\mddefault}{\itdefault}$Y_1$}}}
\put(3451,-6586){\makebox(0,0)[b]{\smash{\SetFigFont{7}{8.4}{\rmdefault}{\mddefault}{\itdefault}$X_1$}}}
\put(3076,-6211){\makebox(0,0)[lb]{\smash{\SetFigFont{8}{9.6}{\rmdefault}{\mddefault}{\itdefault}$A$}}}
\put(2551,-6586){\makebox(0,0)[b]{\smash{\SetFigFont{7}{8.4}{\rmdefault}{\mddefault}{\itdefault}$X_2$}}}
\put(2176,-6211){\makebox(0,0)[lb]{\smash{\SetFigFont{8}{9.6}{\rmdefault}{\mddefault}{\itdefault}$B$}}}
\put(1126,-5086){\makebox(0,0)[rb]{\smash{\SetFigFont{7}{8.4}{\rmdefault}{\mddefault}{\itdefault}$Y_2$}}}
\put(2026,-5086){\makebox(0,0)[rb]{\smash{\SetFigFont{7}{8.4}{\rmdefault}{\mddefault}{\itdefault}$Y_2$}}}
\put(2926,-5086){\makebox(0,0)[rb]{\smash{\SetFigFont{7}{8.4}{\rmdefault}{\mddefault}{\itdefault}$Y_2$}}}
\put(1276,-5311){\makebox(0,0)[lb]{\smash{\SetFigFont{8}{9.6}{\rmdefault}{\mddefault}{\itdefault}$C$}}}
\put(2176,-5311){\makebox(0,0)[lb]{\smash{\SetFigFont{8}{9.6}{\rmdefault}{\mddefault}{\itdefault}$D$}}}
\put(3076,-5311){\makebox(0,0)[lb]{\smash{\SetFigFont{8}{9.6}{\rmdefault}{\mddefault}{\itdefault}$C$}}}
\put(1126,-4186){\makebox(0,0)[rb]{\smash{\SetFigFont{7}{8.4}{\rmdefault}{\mddefault}{\itdefault}$Y_1$}}}
\put(4501,-5911){\makebox(0,0)[b]{\smash{\SetFigFont{8}{9.6}{\rmdefault}{\mddefault}{\itdefault}$X_1$}}}
\put(4501,-4636){\makebox(0,0)[b]{\smash{\SetFigFont{8}{9.6}{\rmdefault}{\mddefault}{\itdefault}$S^\oplus_{11}$}}}
\put(5101,-4636){\makebox(0,0)[b]{\smash{\SetFigFont{8}{9.6}{\rmdefault}{\mddefault}{\itdefault}$S^\oplus_{21}$}}}
\put(5701,-4636){\makebox(0,0)[b]{\smash{\SetFigFont{8}{9.6}{\rmdefault}{\mddefault}{\itdefault}$S^\oplus_{12}$}}}
\put(6301,-4636){\makebox(0,0)[b]{\smash{\SetFigFont{8}{9.6}{\rmdefault}{\mddefault}{\itdefault}$S^\oplus_{22}$}}}
\put(5101,-5911){\makebox(0,0)[b]{\smash{\SetFigFont{8}{9.6}{\rmdefault}{\mddefault}{\itdefault}$Y_1$}}}
\put(6301,-5911){\makebox(0,0)[b]{\smash{\SetFigFont{8}{9.6}{\rmdefault}{\mddefault}{\itdefault}$Y_2$}}}
\put(5701,-5911){\makebox(0,0)[b]{\smash{\SetFigFont{8}{9.6}{\rmdefault}{\mddefault}{\itdefault}$X_2$}}}
\end{picture}